# X-ray Tail in NGC 7619

Dong-Woo Kim[1], Eunhyeuk Kim[2], Giuseppina Fabbiano[1] and Ginevra Trinchieri[3]

1. Smithsonian Astrophysical Observatory, 60 Garden Street, Cambridge, MA 02138
2. Seoul National University, Seoul, South Korea
3. INAF-Osservatorio Astronomico di Brera, Milan, Italy

(Tuesday, July 15, 2008)

## ABSTRACT

We present new observational results of NGC 7619, an elliptical galaxy with a prominent X-ray tail and a dominant member of the Pegasus group. With *Chandra* and *XMM-Newton* observations, we confirm the presence of a long X-ray tail in the SW direction; moreover, we identify for the first time a sharp discontinuity of the X-ray surface brightness in the opposite (NE) side of the galaxy. The density, temperature and pressure jump at the NE discontinuity suggest a Mach number ~1, corresponding to a galaxy velocity of ~500 km s$^{-1}$, relative to the surrounding hot gas. Spectral analysis of these data shows that the Iron abundance of the hot gaseous medium is much higher (1-2 solar) near the center of NGC 7619 and in the tail extending from the core than in the surrounding regions (≲ 1/2 solar), indicating that the gas in the tail is originated from the galaxy. The possible origin of the head-tail structure is either on-going ram-pressure stripping or sloshing. The morphology of the structure is more in line with a ram pressure stripping phenomenon, while the position of NGC 7619 at the center of the Pegasus I group, and its dominance, would prefer sloshing.

*Subject headings*: galaxies: individual (NGC 7619, NGC 7626) – galaxies: ISM – X-rays: galaxies

1. INTRODUCTION

NGC 7619 is one of the two dominant galaxies of the Pegasus I group/cluster (also called U842 or SRGb031) which includes 13 known member galaxies (Ramella et al. 2002). NGC 7619 is also one of the X-ray bright elliptical galaxies observed with *Einstein*, with large hot gaseous halos (Fabbiano, Kim & Trinchieri 1992; Kim, Fabbiano and Trinchieri 1992). The *ROSAT* PSPC observations (Trinchieri, Fabbiano and Kim 1997) confirm that the X-ray emission of NGC 7619 is thermal ($kT \sim 0.8$ keV) and extended as well as asymmetric. The asymmetric core-tail structure of NGC 7619 is stretched to the SW direction several times the length of the opposite emission. The X-ray surface brightness distribution appears to be not only spherically asymmetric but also non-uniform, suggesting dynamically perturbed gas.

Prior to *Chandra*, only a few elliptical galaxies were known to contain asymmetric, disturbed hot halos: NGC 4406 (Rangarajan et al. 1995; Forman et al. 2001), NGC 4472 (Irwin & Sarazin 1996; Biller et al. 2004) and NGC 7619 (Trinchieri, Fabbiano & Kim 1997; this paper). Because they are all in cluster environment and because NGC 4406, in particular, exhibits a large radial (supersonic) motion relative to the cluster, this phenomenon was interpreted as the best evidence of ram pressure stripping (e.g., Forman et al. 1979; White & Sarazin 1991). *Chandra* observations now provide high spatial resolution images so that the fine structural features and physical properties of the compressed and 'stripped' hot ISM can be investigated in detail (e.g., Machacek et al. 2004, 2006).

Another important phenomenon which was identified by Chandra observations is the cold front (e.g., Markevitch et al. 2002; Markevitch & Vikhlinin 2007 and references therein), where the denser gas is colder than the surrounding gas as opposed to the shock front. The cold front is often found near the center of relaxed clusters, where there is no clear sign of recent major mergers such that the X-ray surface brightness distribution would be very smooth without cold fronts. To explain this feature, Ascasibar & Markevitch (2006) introduced a new mechanism, called sloshing, which may explain the observational features of the cold fronts without any other obvious disturbance (as opposed to originally suggested mergers by Markevitch et al. 2002; see Markevitch & Vikhlinin 2007 for more details).

In this paper, we present the results of X-ray observations of NGC 7619 with *Chandra* and *XMM-Newton*, resulting in high resolution images for a detailed study of the spatial structures, and high S/N spectra for determining the physical properties of the hot ISM, particularly at the head and tail. Table 1 provides a complete list of all the X-ray observations of NGC 7619 with observation dates, exposure times and numbers of point sources detected.

Table 1
X-ray Observations of NGC 7619

| Mission | obs id | obs date | exp (ksec) | # point src | reference |
|---|---|---|---|---|---|
| EINSTEIN IPC | 2598 | Sep. 11 1980 | 9 | | Fabbiano, Kim & Trinchieri (1992) |
| ROSAT PSPC | 600134 | May 30, 1992 | 18 | 26 | Trinchieri, Fabbiano & Kim (1997) |
| ROSAT HRI | 600942 | Jun. 8, 1997 | 8 | <10 | -- |
| Chandra ACIS-I | 02074 | Aug. 20, 2001 | 27 | 92 | -- |
| Chandra ACIS-S | 03955 | Sep. 24, 2003 | 25 | 78 | this paper |
| XMM-Newton | 0149240101 | Dec. 16, 2003 | 40 | 96 | this paper |

This paper is organized as follows: In section 2, we describe the *Chandra* and *XMM-Newton* observations and basic data reductions. In section 3, we present our results of the spatial analysis, particularly for the head-tail structures. In section 4, we present our results of the spectral analysis, emphasizing the metal abundances at various regions. In section 5, we discuss possible origins of the head-tail structure and their implications. Finally we summarize our conclusions in section 6.

Throughout this paper, we adopt a distance to NGC 7619 D=52.97 Mpc, based on the surface brightness fluctuation analysis by Tonry et al. (2001). At the adopted distance, 1′ corresponds to 15.4 kpc.

2. X-RAY OBSERVATIONS

2.1. *Chandra* Observations

NGC 7619 was observed for 40 ksec on September 24, 2003 with the *Chandra* Advanced CCD Imaging Spectrometer (ACIS; Weisskopf et al. 2000) (obsid = 3955). The *Chandra* ACIS field of view is overlaid on the optical DSS image in Figure 1 (red squares). The ACIS data were reduced in a similar manner as described in Kim & Fabbiano (2003) with a custom-made pipeline (XPIPE), specifically developed for the *Chandra* Multi-wavelength Project (ChaMP; Kim et al. 2004). To apply up-to-date calibration data (e.g., CCD gain, bad pixel map), we regenerated new level 2 data products by re-running ***acis_process_event*** (see http://asc.harvard.edu/ciao/guide/ acis_data.html). The *Chandra* observations suffered from significant background flares which are most significant in CCD S3. Removal of background flares reduced the effective exposure time of CCD S3 from 37.5 ksec to 24.5 ksec. We detected X-ray point sources with ***wavdetect*** (Freeman & Vinay 2002) available in CIAO (see http://asc.harvard.edu/ ciao) and applied on these detection aperture photometry with the 95% encircled energy circle determined at 1.5 keV (http://cxc.harvard.edu/ cal/ Hrma/ psf) to extract source properties (Kim et al. 2004). NGC 7619 was previously observed with ACIS-I CCDs (obsid=2074) for 30 ksec. This field of view is also marked in Figure 1 (green squares). Since the ACIS-S back-illuminated CCD is more sensitive to the soft X-rays (< 1.5 keV) than the front-illuminated ACIS-I CCDs and the ACIS-I field of view does not fully cover the extended tail region, we primarily use the new ACIS-S data to investigate the X-ray tail. We use ACIS-I data to study objects (e.g., NGC 7626) not covered by ACIS-S and to check the consistency between different data.

2.2. *XMM-Newton* Observations

NGC 7619 was observed for 40 ksec on December 16, 2003 (obsid = 0149240101) with *XMM-Newton* (Jansen et al. 2001). The XMM MOS field of view is overlaid on the optical DSS image in Figure 1 (big blue circle). The XMM data were reduced with SAS version 6.0 (http://xmm.vilspa.esa.es/sas) and with the prescription given by Snowden et al. (2004). To remove the background flare, we generate a light curve for each instrument at 10-15 keV (and pattern=0). After screening the flares, the effective exposure becomes 39.5 ksec, 39.8 ks and 36.7 ks for MOS1, MOS2 and PN, respectively. For both spatial and spectral analyses of the extended diffuse hot ISM, it is critical to determine the background contribution accurately. This is particularly important for the XMM data because of the larger area coverage and higher background contribution. Therefore, we utilized the blank sky background (Read and Ponman 2003) obtained from http://xmm.vilspa.esa.es/external/xmm_sw_cal/background, and we apply the double background subtraction technique described by Arnaud et al. (2002). This method removes both the X-ray and non X-ray components of the background and allows accurate extraction of source spectra and surface brightness profiles. The blank sky backgrounds were screened for flares using the same criteria (i.e., the same count rate at the same energy band) as in our observations. Applying the SAS *evigweight* tool, we corrected photon by photon both observations and the blank sky backgrounds for telescope vignetting. We then subtracted the sky background after scaling by the ratio of count rates determined at 10-12 keV for MOS and 12-14 keV for PN; the scale factors are only ~10% for all three instruments, because both observation and blank sky data were screened in the same way. We finally subtracted the residual background determined from a source free region, at r > 10′ and away from the extended tail (see section 3.2). This region may still have a low level of the remaining diffuse emission, but the residual from the scaled blank sky background was almost negligible in the soft energy (< 2.5 keV).

3. SPATIAL ANALYSIS

3.1 Point-like X-ray Sources

We detect 78 point-like sources in the *Chandra* ACIS-S observation and 32 sources in the S3 chip only (Figure 2). Given the distance of NGC 7619 and the presence of strong diffuse emission, a typical LMXB (with $L_X < 10^{39}$ erg s$^{-1}$) is not easily detected. We found only 1 non-nuclear point source inside the $D_{25}$ ellipse of NGC 7619. Its position is at (RA, Dec) = (23 20 11.9, 8 11 25), 70″ SW from the nucleus of NGC 7619. If it is associated with NGC 7619, the X-ray luminosity of this source, $L_X = 5 \times 10^{39}$ erg s$^{-1}$ makes it a ULX candidate. The chance probability of being a background AGN is 25%, based on the ChaMP log(N)-log(S) relationship of the cosmic background X-ray sources (Kim M. et al 2007). However, its soft X-ray emission may suggest it is not a background AGN (see section 4.5). We detect 96 sources in the *XMM-Newton* observation from the combined MOS 1+2 image (see Figure 3). Again, inside the $D_{25}$ ellipse we detect only one source, as seen in the *Chandra* observation.

We also detect X-ray emission from other galaxies in our field of view (see Figure 1): NGC 7617 (2.8′ SW from NGC 7619), NGC 7626 (6.9′ E of NGC 7619), NGC 7623 (12′ NW from NGC 7619) and NGC 7611 (12.7′ SW from NGC 7619).

We list the optical and X-ray positions of these galaxies and the ULX candidate in Table 2. None of them are resolved in the ACIS observations, except NGC 7626. The results of spectral analysis of these galaxies are presented in section 4.5.

Table 2
Optical and X-ray positions of other galaxies and the ULX candidate

| name | optical position RA DEC (J200) | X-ray position RA DEC (J2000) | D from NGC 7619 ('/kpc) | B (mag) | T |
|---|---|---|---|---|---|
| NGC 7617 | 23 20  9.0   8  9 57 | 23 20  9.0   8  9 57 | 2.8/ 43 | 14.5 | -2 |
| NGC 7626 | 23 20 42.5   8 13  1 | 23 20 42.5   8 13  1 | 6.9/106 | 12.0 | -5 |
| NGC 7623 | 23 20 30.0   8 23 45 | 23 20 29.9   8 23 46 | 12.0/185 | 13.6 | -1.5 |
| NGC 7611 | 23 19 36.6   8  3 48 | 23 19 36.5   8  3 49 | 12.7/196 | 13.4 | -1 |
| ULX | -- -- | 23 20 11.9   8 11 25 | 1.2/ 18 | -- | -- |

The optical position are from NED for galaxies. The X-ray positions are from the *Chandra* ACIS-S observation, except NGC 7626 and NGC 7623. For NGC 7626 we used the *Chandra* ACIS-I observation (it falls outside of the ACIS-S fov). For NGC 7623, we used the XMM MOS data (its PN position is 3″ off; it is within the ACIS-I fov, but not detected.) The optical magnitude and morphological type are from RC3.

About 70% of unresolved point-like sources are detected in the area covered by the X-ray tail. As seen in Figure 2, 23 out of 32 X-ray point sources detected in the ACIS S3 chip are found in the S-W direction from NGC 7619 with PA=170-260°. Taking into account the effective area of 45% of the whole S3 chip (determined with the exposure map), we estimate the significance of the excess number of sources in the tail region (over the other sources found in the remaining region of the S3 chip) to be 2.7σ. We also compare the number of these point sources with the number of expected X-ray background sources, determined from the ChaMP logN-logS relation (Kim, M. et al. 2007). As plotted in Figure 4, we find that the statistical significance of the excess number of sources is at 3σ level in both soft and hard bands at $F_X \sim 10^{-15}$ erg cm$^{-2}$ s$^{-1}$ in the soft band (0.5-2 keV), or $F_X \sim$ a few $10^{-15}$ erg cm$^{-2}$ s$^{-1}$ in the hard band (2-8 keV). Some of these sources have soft X-ray spectra (only detected in the soft energy band), reminiscent of those possibly cooler blobs seen at the periphery of the X-ray emitting halo of NGC 507 (Kim and Fabbiano 1995; 2004). We will discuss these sources in a future paper; here we will concentrate on the diffuse X-ray emission from the hot ISM of NGC 7619.

3.2 Diffuse X-ray Emission

Figure 2 and 3 show the images obtained from the soft band (0.3-2.5 keV) *Chandra* ACIS S3 and *XMM-Newton* (MOS 1+2 combined) data, smoothed with Gaussians with sigma of 5″ and 7.5″, respectively. To emphasize the diffuse emission, we also show in Figure 5 the ACIS S3 image obtained in a narrow energy band (0.7 – 1.2 keV) after excluding all the detected point sources from the data (based on the PSF at the source location), applying exposure correction to eliminate instrumental effects, and smoothing with a 12" Gaussian. The head-tail structure is clearly seen: toward the NE (i.e., head-side) there is a clear discontinuity in the surface brightness at r = 1.2 - 1.5′ (18 - 23 kpc from the galaxy nucleus), while toward the SW (i.e., tail-side) the X-ray emission is extended to r ~ 15′, or 230 kpc (see the *XMM-Newton* image in Figure 3 and the radial profile in Figure 7). The elongated extended feature appears to form two high-surface brightness tails with a lower surface brightness gap in between (see Figure 5). The statistical significance of the brightness difference between the tails and the gap is ~2σ and ~3σ (4σ) determined from the ACIS and MOS1+2 (MOS +PN) images, respectively. We call these two tails: main tail (T1) at PA ~ 190° and secondary tail (T2) at PA ~ 240°.

To describe more quantitatively the head-tail structure, we generated radial profiles of the X-ray surface brightness in different azimuthal sectors. The leading edge is most clearly seen at PA=-20° – 100° while the tail is more prominent at PA= 170° – 260° (see Figure 6). In Figure 7, we compare the surface brightness distributions toward the head (red) and tail (black) directions made with the ACIS S3 (filled) and the XMM MOS 1+2 images (open). The radial slopes of ACIS and MOS surface brightness are consistent with each other within the statistical errors, after normalized for different effective areas. In both ACIS and MOS radial profiles, it is obvious that the X-ray emission is more prominent toward the tail direction. Toward the head direction, the diffuse emission is extended to ~10′, while toward the tail it is extended to 15′-20′, i.e., to the edge of the detector, as previously reported with the ROSAT data (Trinchieri et al. 1997). Toward the leading edge, the X-ray emission drops abruptly at r = 1.2′, after a local flattening ('bump'). The red solid line is the best fit power-law model (slope = 1.68 ± 0.04) determined for the head side radial profile excluding the bump near the discontinuity (r= 0.7′-1.3′) and the central region (r < 2″). The black solid line is a power-law (with the same slope as in the head side) with an additional $\beta$-model for the extended tail; the best fit model parameters are $\beta$ = 0.41 ± 0.07, corresponding to a power-law slope of 1.48 ± 0.4 at large galacto-centric distances and the core radius, $r_c$ = 15.8 (-4 +33). To better show the bump at r~1′, we expand this part of the surface brightness profile in Figure 8 where the radius is in a linear scale. The surface brightness enhancement at the bump may indicate a shell-like feature (see Figure 5) caused by ram-pressure resulting from the motion of NGC 7619 in the NE direction (see Section 5).

## 4. SPECTRAL ANALYSIS

To extract X-ray spectra from various regions of the images, we use CIAO *dmextract* for *Chandra* data and SAS *xmmselect* for *XMM-Newton* data. Each spectrum is then binned to have at least 25 counts per bin in order to properly perform a $\chi^2$ fit. We limit the spectral fitting to the energy range of 0.5 – 5 keV for both *Chandra* and *XMM-Newton* spectra to avoid the significantly dominating background at high energies (> 5keV) and the calibration uncertainty at lower energies (< 0.5 keV). For fitting *XMM-Newton* spectra, we have also tried excluding the narrow energy range 1.37 – 1.6 keV to remove the effect of the strong Al-K instrument line. However we did not find any significant difference in our results, within the statistical error. For each spectrum, we determined a redistribution matrix file (RMF) and an auxiliary response file (ARF) for each source region using CIAO and SAS tools.

To determine the X-ray spectral properties of the hot ISM, it is important to apply a realistic emission model, consisting of multiple emission components. The model dependence is most critical in measuring metal abundances, which has long been a controversial subject (e.g., Kim & Fabbiano 2004 and reference therein). Although the reality is likely to be more complex, we take a three-component model to represent the low-T (<0.8 keV) ISM gas (**MEKAL** or **VMEKAL** model) in the galaxy, the high-T (>1.0 keV) group ambient gas (another **MEKAL** or **VMEKAL**) and a hard (5-10 keV) X-ray component (**BREM**) to account for undetected LMXBs and background AGNs. For the hard emission, we fix the temperature to be 7 keV (Irwin et al. 2003; Kim & Fabbiano 2004). We note that the hard component may also be represented by a power-law of a photon index of ~1.7 and the results are consistent within the error. The relative normalizations of the three components are free to vary to reflect possible different contributions to the spectrum arising from these different components in different spatial regions.

### 4.1 Result for concentric annuli

In Table 3, we summarize the goodness of fit for the spectra extracted from circular annuli (0′-1′, 1′-2′, 2′-3′, and 3′-5′) from different instruments, for $N_H$ both fixed at the line of sight value (5 x $10^{20}$ cm$^{-2}$), and left free to vary during the fit. We fit individual spectra extracted from each instrument as well as jointly fit multiple spectra with the same set of spectral parameters. In most cases, the goodness of fit is acceptable with $\chi^2_{red}$ close to one (< 1.1) and the results of different combinations of spectra are consistent with each other within the statistical error. However, the joint fit of MOS and PN spectra results in a considerably higher $\chi^2_{red}$ (~1.4 for ~1600 degree of freedom), although the parameters are still consistent. Therefore, in the following we present the fitting results of MOS (MOS1 + MOS2 combined) and PN spectra, separately. We determine de-projected quantities by fitting 2D spectra to the projected 3D models using **xspec project**. (heasarc.gsfc.nasa. gov/ heasoft/ xanadu/ xspec).

TABLE 3

Goodness of Fit for different instruments

```
---------------------------------
Instrument   reduced_chi2(Chi2/dof)
---------------------------------
[N(H) fixed, deprojected]
 mos12        1.12( 794/ 712)
  mos1        1.09( 379/ 346)
  mos2        1.15( 392/ 342)
    pn        1.04( 955/ 919)
 mospn        1.41(2342/1655)

[N(H) free, deprojected]
 mos12        1.06( 753/ 708)
  mos1        1.07( 365/ 342)
  mos2        1.08( 364/ 338)
    pn        1.03( 946/ 915)
 mospn        1.39(2297/1651)
---------------------------------
```

We list the best fit parameters and their errors in Table 4. Throughout this paper, we quote errors determined at 90% confidence for 1 significant parameter. If the upper or lower limit (or sometimes both) is not statistically constrained, the limit remains blank. Also listed are the individual fluxes (in unit of $10^{-13}$ erg cm$^{-2}$ s$^{-1}$) of three emission components to show their relative importance in the different radial bins. When $N_H$ is set to vary, its best fit value is slightly higher than the Galactic line of sight value (5.0 x $10^{20}$ cm$^{-2}$) in the central bin (1.2 x $10^{21}$ cm$^{-2}$ for MOS and 9 x $10^{20}$ cm$^{-2}$ for PN) and it becomes lower (2 x$10^{20}$ cm$^{-2}$) than the Galactic value in the outermost annulus (r=3′-5′). Given that there is no absorption reported at other wavelengths (see section 5) and that $N_H$ cannot go below the Galactic line of sight value, we fix $N_H$ to be at the Galactic value (see below for the effect of $N_H$ on the measured metal abundance). The temperature ($kT_1$) of the hot ISM is 0.7 keV at the center and remains almost constant to r = 3′ (or slightly increases to 0.8 keV). In the outermost annulus (r = 3′-5′), which is outside the $D_{25}$ ellipse, $kT_1$ is not well determined. The temperature ($kT_2$) of the ambient gas is always near ~1.1 keV in the outer regions, but it is not well constrained in the center. The individual fluxes of three emission components (Table 4) indicate that the X-ray emission is dominated by the 0.7 keV hot ISM near the center, while the ambient gas ($kT_2$ ~1.1 keV) dominates at the outskirts (i.e., outside the $D_{25}$ ellipse). A similar trend was seen in NGC 507 (Kim & Fabbiano 2004), another elliptical galaxy in a group. We take $kT_2$ = 1.1 keV as a temperature of the ambient gas in the Pegasus I group (see section 5). We also confirmed the temperature using the spectra extracted from the regions near, but outside the tail region (see below in this section). This is consistent with the previous *ROSAT* estimate ($1.27 ^{+0.6}_{-0.2} keV$) of Trinchieri et al. (1997), within the error. In Table 5, we summarize the fitting results with $kT_2$ fixed at 1.1 keV.

### TABLE 4
Fitting Result with Spectra Extracted from Concentric Annuli

```
[N(H) fixed, deprojected]
```

```
--------------------------------------------------------------------------------
              r=0'-1'              r=1'-2'              r=2'-3'              r=3'-5'

                                        MOS1+MOS2
--------------------------------------------------------------------------------
Z_Fe     1.65 (1.24-2.50)  1.20 (0.70-2.55)  2.61 (0.58-    )  0.79 (0.41-1.89)
kT1      0.71 (0.70-0.72)  0.79 (0.68-0.82)  0.68 (0.51-0.86)  0.11 (    -    )
kT2      2.00 (    -    )  1.00 (    -    )  1.00 (    -    )  1.05 (1.01-1.07)

Fx1      5.14 (2.57-6.73)  1.99 (0.87-3.35)  0.39 (0.17-3.32)  0.11 (    -0.28)
Fx2      0.41 (    -0.70)  0.66 (    -    )  1.34 (    -5.83)  3.03 (1.56-4.64)
Fx3      0.97 (0.53-1.45)  0.94 (0.12-1.66)  1.27 (    -3.06)  3.27 (2.63-3.98)
--------------------------------------------------------------------------------

                                           PN
--------------------------------------------------------------------------------
Z_Fe     2.94 (1.77-4.99)  3.15 (1.24-    )  0.97 (0.49-2.51)  0.72 (0.49-1.15)
kT1      0.69 (0.68-0.71)  0.82 (0.70-0.85)  0.74 (0.62-0.86)  0.69 (0.48-0.87)
kT2      1.47 (    -    )  2.00 (    -    )  1.08 (    -    )  1.15 (1.08-1.33)

Fx1      3.79 (2.12-6.94)  2.53 (1.08-5.64)  1.16 (0.34-2.73)  0.77 (0.18-1.97)
Fx2      0.34 (    -    )  1.13 (    -2.84)  1.83 (    -5.16)  4.35 (3.15-5.51)
Fx3      0.44 (    -0.87)  0.85 (0.06-1.72)  0.79 (    -    )  2.39 (1.00-3.27)
--------------------------------------------------------------------------------

Z_Fe : Fe abundance in solar unit (taken from Grevesse, N. & Sauval, A.J. 1998).
       All elements are varied together at the solar abundance ratio.
kT1  : Temperature of the 1st MEKAL emission component (< 1 keV)
kT2  : Temperature of the 2nd MEKAL emission component (1-2 keV)
kT3  : fixed at 7 keV for Brem
Fx1  : flux (0.3-8.0 keV) from the 1st MEKAL component
Fx2  : flux (0.3-8.0 keV) from the 2nd MEKAL component
Fx3  : flux (0.3-8.0 keV) from the 7 keV Brem component
```

Following Grevesse & Sauval (1998), we first set all elements to vary together at the solar ratio (see below for fitting with variable abundance ratios). With these constraints, we find that the metal abundance (mainly driven by the *Fe* abundance) gradually declines with increasing radius. While the *Fe* abundance is super-solar in the central region of NGC 7619, it becomes sub-solar outside the D25 ellipse (r > 2′). If $N_H$ is fixed to the Galactic value and $kT_2$ is fixed at 1.1 keV, $Z_{Fe}$ is 1.6 (1.2-2.2) $Z_\odot$, or 2.6 (1.8 -4.8) $Z_\odot$ for MOS and PN spectral fitting, respectively. This super-solar abundance is similar to that of giant elliptical galaxies, measured with a similar method from XMM observations (e.g., NGC 5044 Buote et al. 2003 and NGC 507 Kim & Fabbiano 2004). These super-solar abundances are consistent with what is expected from the accumulation of SN synthesized metals in elliptical galaxies (e.g., Arimoto et al. 1997), and exceed previous reports of metal abundances in elliptical halos (typically using only one or two-component models in the fit; see e.g., Awaki et al. 1994). At the outskirts, instead, we find a lower $Z_{Fe}$ (0.4-0.9 $Z_\odot$).

While the spectral fitting with the projected 3D 3-component model is most suitable for the X-ray emission from the ISM in the galaxy, this may be over-modeled for the ICM X-ray emission in the outskirts. Because the ambient gas outside the head-tail structure is likely in a single temperature (or slowly varying in space) and because LMXBs and ISM do not contribute much, we can determine the ambient gas properties by applying a single emission model. We re-extract the spectra from the regions just outside the X-ray tail, at the edge (lower-left and upper-right corners) of the central chip (ccdid=1) of MOS 1 and MOS 2 (see Fig 3). Then, we apply a single component **MEKAL** model. The best fit parameters are more tightly constrained to be $kT = 1.1 ^{+0.2}_{-0.1} keV$ and $Z_{Fe} = 0.3 ^{+0.2}_{-0.1} Z_\odot$ with the reduced $\chi^2 = 1.0$ with 120 degrees of freedom. The *Fe* abundance is consistent with the typical abundance found in the ICM, 0.3-0.5 $Z_\odot$ (e.g., Mushotzky et al. 1996; Fukazawa et al. 1998).

In summary, the ISM inside the $D_{25}$ ellipse of NGC 7619 is maintained at the temperature of $kT = 0.7$-$0.8$ keV with a super solar metal abundance, while the ambient gas in the outskirts of NGC 7619 is hotter (1.1 keV) and metal-poorer (0.3-0.5 $Z_\odot$) than the ISM.

TABLE 5
same as Table 4, but $kT_2$ fixed at 1.1 keV

```
[N(H) fixed, deprojected, kT2 fixed]
-------------------------------------------------------------------------------
            r=0'-1'          r=1'-2'          r=2'-3'          r=3'-5'

                                    MOS1+MOS2
-------------------------------------------------------------------------------
Z_Fe    1.61 (1.18-2.22)  1.06 (0.65-2.36)  2.58 (0.82-    )  0.61 (0.43-0.99)
kT1     0.71 (0.70-0.72)  0.80 (0.71-0.83)  0.78 (0.62-0.87)  0.95 (0.56-    )

Fx1     5.28 (3.38-7.04)  2.26 (0.77-2.90)  0.88 (0.23-5.01)  0.97 (0.31-2.95)
Fx2     0.13 (    -    )  0.42 (    -    )  0.88 (    -8.63)  2.28 (0.42-3.75)
Fx3     1.15 (0.89-1.41)  0.88 (    -1.67)  1.22 (    -2.99)  3.03 (2.26-3.78)
-------------------------------------------------------------------------------

                                       PN
-------------------------------------------------------------------------------
Z_Fe    2.55 (1.82-4.84)  3.37 (2.42-    )  1.11 (0.56-2.19)  0.68 (0.46-0.83)
kT1     0.69 (0.68-0.71)  0.81 (0.71-0.85)  0.71 (0.64-0.84)  0.56 (0.36-0.79)

Fx1     3.75 (1.98-5.54)  2.33 (1.05-3.98)  0.91 (0.27-1.62)  0.35 (0.06-1.06)
Fx2     0.35 (    -    )  1.16 (    -4.26)  2.07 (    -3.24)  4.69 (3.28-6.25)
Fx3     0.48 (    -1.20)  1.07 (0.56-1.59)  0.82 (    -    )  2.48 (1.35-4.07)
-------------------------------------------------------------------------------
```

Since the iso-intensity contour of the X-ray surface brightness is elliptical rather than circular (see Figure 5), we have also extracted the spectra in elliptical annuli (blue ellipses in Figure 6) by keeping the semi-major axes the same as radii in the circular annuli in the above, but with a/b=3/2 and PA of the major axis=-50° measured from the north. Repeating the spectral fitting by the same way described above, we do not find any statistically significant difference from the results reported in Table 4 and 5.

4.2 The tails

We then extracted X-ray spectra from the tail (after excluding point sources) to determine its physical properties. If the hot gas in the tail is indeed originated from the galaxy, we expect the tail to have temperature and metal content similar to those of the ISM, rather than the ICM. Since the extended gas may consist of two tails, (section 3.2), we extracted the X-ray spectra in two elliptical regions (marked by red colored ellipses in Figure 6). For the fit, we fixed the temperature of the ambient gas to be $kT_2$ = 1.1 keV (as determined in the above) and $N_H$ to be the Galactic value. The best fit $Z_{Fe}$ is 1 – 2 solar and $kT_1$ is 0.8 ± 0.1 keV, for both tails with the main tail being better constrained with a higher X-ray flux (Table 6). While the physical properties of the main tail (T1) and the hot ISM inside the $D_{25}$ ellipse are identical within the error, the difference in $kT$ ($Z_{Fe}$) between T1 and the surrounding gas (section 4.1) is 2.5σ (2.8σ). We also fixed the abundance to be 0.5 solar for the 2$^{nd}$ **MEKAL** component (i.e., the ambient gas as determined above), but the results do not change significantly, because the X-ray emission is dominated by the first **MEKAL** component. The tail is clearly cooler than the ICM at a similar radial distance from the center of NGC 7619 and metal-enriched with a super-solar metal abundance, strongly indicating that the gas in the tail is indeed originating from the galaxy.

TABLE 6
Fitting Result with Spectra Extracted from the X-ray Tail

```
[N(H) fixed, not deprojected, kT2 fixed]
-------------------------------------------------
                T1                      T2
                        MOS1+MOS2
-------------------------------------------------
Z_Fe      1.19 (0.99-2.26)       1.03 (0.61-1.71)
kT1       0.82 (0.70-0.87)       0.95 (0.79-    )

Fx1       0.61 (0.11-1.30)       0.52 (0.15-8.44)
Fx2       0.10 (    -0.44)       0.12 (    -0.59)
Fx3       0.52 (0.37-0.68)       0.29 (0.12-0.41)
-------------------------------------------------

                        PN
-------------------------------------------------
Z_Fe      1.26 (0.81-2.67)       0.97 (0.65-1.58)
kT1       0.82 (0.70-0.88)       1.00 (0.78-    )

Fx1       0.91 (0.30-1.25)       0.85 (0.11-1.38)
Fx2       0.44 (    -0.71)       0.10 (    -7.62)
Fx3       0.45 (0.20-0.73)       0.38 (0.14-0.58)
-------------------------------------------------
```

Note. The two tail regions, T1 (lower-left) and T2 (upper-right) are marked by red ellipses in Figure 6.

4.3 α-to-*Fe* abundances

We also measure α-elements, using a **VMEKAL** emission model (Table 7). Due to the limited statistics, we set elements lighter than *Ca* to vary with *Si* and the other elements to vary with *Fe*. In all radial bins, the α-to-*Fe* abundance ratio is consistent with the solar ratio within the statistical error, although the α-elements appear to be slightly underabundant compared to *Fe*, with the best-fit ratio being close to (~0.8). The α-to-*Fe* abundance ratio being close to (or slightly lower than) solar is also seen in NGC 507 (KF04) and NGC 1316 (Kim et al. in prep). This is contrary to the stellar metal abundance measured by the optical observations where typical giant elliptical galaxies tend to be α-elements enriched (e.g., Trager et al. 2000). The α-to-*Fe* abundance ratio in the tail is not well constrained.

TABLE 7
Fitting Result with Spectra Extracted from Concentric Annuli

```
[N(H) fixed, deprojected, kT2 fixed]
--------------------------------------------------------------------------------
            r=0'-1'            r=1'-2'             r=2'-3'            r=3'-5'

                                      MOS1+MOS2
--------------------------------------------------------------------------------
Z_Fe    1.40 (1.03-2.15)   0.92 (0.59-1.80)   3.58 (0.51-    )   0.57 (0.43-0.81)
Z_α     1.22 (0.77-2.12)   0.65 (0.23-1.94)   5.00 (0.79-    )   0.10 (    -0.35)
kT1     0.71 (0.70-0.73)   0.79 (0.70-0.83)   0.80 (0.61-0.87)   1.00 (    -    )

Fx1     5.10 (3.39-6.86)   2.12 (0.82-3.45)   0.91 (0.21-6.04)   0.72 (    -1.89)
Fx2     0.27 (    -1.28)   0.51 (    -2.60)   0.77 (    -    )   2.21 (1.04-3.57)
Fx3     1.19 (0.87-1.47)   0.97 (0.03-1.97)   1.32 (    -3.87)   3.48 (3.01-4.23)
--------------------------------------------------------------------------------

                                         PN
--------------------------------------------------------------------------------
Z_Fe    1.79 (1.14-3.78)   1.66 (0.86-    )   0.82 (0.44-1.76)   0.61 (0.47-0.93)
Z_α     1.31 (0.66-3.72)   0.65 (    -3.82)   0.34 (    -1.91)   0.17 (    -0.49)
kT1     0.69 (0.66-0.70)   0.80 (0.70-0.86)   0.70 (0.56-0.85)   0.53 (    -    )

Fx1     3.43 (1.38-5.33)   1.84 (0.56-4.34)   0.69 (0.27-1.90)   0.23 (    -0.47)
Fx2     0.58 (    -2.12)   1.46 (    -4.79)   2.09 (0.68-3.85)   4.29 (2.91-5.82)
Fx3     0.63 (    -1.49)   1.41 (0.62-2.05)   1.10 (    -    )   3.35 (2.07-4.38)
--------------------------------------------------------------------------------

Note. Same as Table 5, but using VMEKAL to separately measure Fe and α-element
abundances
```

4.4 The leading edge

To determine the physical properties across the discontinuity along the leading edge, we extracted the X-ray spectra in concentric conic annuli (PA=-20° – 100°) from both XMM and *Chandra* data. We applied a two-component model (**MEKAL** and the 7 keV Bremsstrahlung) to measure the emission-weighted average gas temperature for a given radius. We also used a single **MEKAL** model outside of the edge (since the hard component is not significant there), but the results are consistent within the error. In Figure 9, we plot the temperature of the **MEKAL** component against radius. At the outskirts (r > 2′), the temperature is ~1.1 keV as seen in the above analysis. Across the discontinuity, the temperature drops inward from 1.1 keV to 0.7 keV, as opposed to the expected behavior in a shock front, but consistent with the typical cold front (e.g., Markevitch and Vikhlinin 2007). It is interesting to note that the temperature change only occurs at r = 1 − 2′, at or just outside the discontinuity and the temperature is nearly constant (either at 0.7 or 1.1 keV) at all other radii. While the *XMM-Newton* spectra (circle and diamond) seems to show a smoother temperature change, the *Chandra* spectra (triangle) suggests a rather abrupt transition between 0.7 and 1.1 keV. Although the ACIS data has a lower S/N, its higher spatial resolution shows a sharp transition at r=2′ with a width of δr ≤ 0.5′ (or ~8 kpc). A deeper *Chandra* observation will be able to determine the exact location and depth of the transition zone. The metal abundance and its variation across the discontinuity are not well constrained, but overall consistent with solar values (with a large error).

4.5 NGC 7626 and other galaxies

NGC 7626, another dominant galaxy in the Pegasus I group, was previously detected in Einstein (Fabbiano et al. 1992) and ROSAT (Trinchieri et al, 1997) observations. Using the ACIS-I and XMM-Newton data (it falls outside of the ACIS-S fov), we extract X-ray spectra from r < 0.5′ and r < 1.5′, respectively. Although they do not fit to an absorbed single emission component model, all spectra (ACIS-I, MOS1+2, PN) fit well (reduced $\chi^2$ close to 1-1.4) to two-component model (**MEKAL** + **power-law**, or **MEKAL** + **BREM**) which represents the soft X-ray emission from the hot ISM and the hard emission from AGN+LMXBs (see Table 8). For the hard component, we fixed power-law photon index $\Gamma_{ph}$=1.7 (or $kT$ = 7 keV for **BREM**). The best fit gas temperature is 0.66 ± 0.03 keV, close to that of NGC 7619. If we allow the metal abundance to vary, the best fit abundance is 0.8 ± 0.4 solar. We also added a 1.1 keV **MEKAL** component to represent the ambient gas found at the outskirts of NGC 7619, but the parameters do not change much. The total X-ray flux in 0.3 – 8.0 keV, after correcting for absorption, is 4 - 6 x $10^{-13}$ erg s$^{-1}$ cm$^{-2}$ (the soft/hard components contribute about 60/40%).

We have also extracted spectra from other galaxies, NGC 7611, NGC 7617 and NGC 7623 and the ULX candidate (see section 2) and performed the same spectral analysis. Given that their X-ray fluxes are lower (by a factor of ~10) than that of NGC 7626, we can only marginally constrain the parameters. The X-ray spectrum of NGC 7611 fits to a single power-law with $\Gamma_{ph}$=1.7 ( 1.0 (close to that of a typical AGN and LMXBs) and it does not require any thermal gas emission, indicating that NGC 7611 contains no or little hot gas. On the other hand, NGC 7617 and 7623 seem to have an additional gas component, although statistically not required. The best fit parameter in a single power-law fit is too steep with (ph=3-4 (i.e., very soft) for a typical AGN and LMXBs. In a two-component model fit (MEKAL + power-law with fixed (ph=1.7 for the hard component), the soft component has 0.7 and 0.2 keV for NGC 7617 and NGC 7623, respectively. In both galaxies, the soft and hard components contribute equally within the error. Given the low statistics, we can not constrain the metal abundance in any of these three galaxies. The ULX candidate is similar to NGC 7617 with (ph=2-3 in a single power-law model and kT=0.8-1 keV in a two-component model with fixed (ph=1.7 for the hard component. This seems to be softer than typical AGN spectra (either absorbed or unabsorbed), possibly suggesting that this X-ray source is associated with NGC 7619.

TABLE 8
Spectral Fitting Results of NGC 7626 and other galaxies

```
wabs*(power-law)    [N(H)=5e20]
--------------------------------------------------------------------------------
           N7626          N7611         N7617         N7623          ULX

                                    MOS1+MOS2
--------------------------------------------------------------------------------
alpha      2.5  ( - )    1.7  (1.0)   3.1  (1.0)    4.4  (2.5)    2.0  (0.4)
Fx         5.0  ( - )    0.19 (0.1)   0.31 (0.06)   0.34 (0.2)    0.13 (0.03)
ch2/dof    477.09/129    5.22/3       3.39/2        2.74/4        0.91/2

                                       PN
--------------------------------------------------------------------------------
alpha      2.6  ( - )    1.3  (0.6)   2.7  ( - )    3.5  (1.6)    3.1  (0.6)
Fx         8.4  ( - )    0.36 (0.1)   0.38 ( - )    0.22 (0.1)    0.36 (0.06)
ch2/dof    833.87/189    7.61/6       21.41/7       8.08/9        3.91/3
--------------------------------------------------------------------------------

wabs*(mekal + power-law)   [Photon_index=1.7 & Z_Fe=solar & N(H)=5e20]
--------------------------------------------------------------------------------
           N7626          N7611         N7617         N7623          ULX

                                    MOS1+MOS2
--------------------------------------------------------------------------------
kT         0.66 (0.03)   0.2  ( - )   0.7  (0.4)    0.2  (0.3)    1.1  ( - )
Fx         4.4  (0.3)    0.41 ( - )   0.20 (0.15)   0.18 (0.1)    0.11 ( - )
ch2/dof    169.70/128    4.20/2       1.38/1        1.98/3        0.89/1

                                       PN
--------------------------------------------------------------------------------
kT         0.65 (0.02)   -    ( - )   0.7  (0.1)    0.4  (0.3)    0.8  (0.2)
Fx         6.2  (1.1)    0.35 ( - )   0.29 (0.1)    0.13 (0.1)    0.24 (0.1)
ch2/dof    212.88/188    7.63/5       4.58/6        7.97/8        2.04/2
--------------------------------------------------------------------------------

Fx (0.3-8.0 keV) in unit of 10^-13 erg/s/cm2.
```

## 5. DISCUSSION

### 5.1 Head-tail structure

Analyzing both *Chandra* and *XMM-Newton* observations, we confirm the existence of an X-ray tail in NGC 7619, which was previously reported with *Einstein* (Fabbiano et al. 1992) and *ROSAT* (Trinchieri et al. 1997) data. We also identify a significant discontinuity in the opposite direction. The discontinuity suggests that the ISM in NGC 7619 is experiencing ram pressure from the NE direction (PA ~ 40°). To determine the physical status of the ISM relative to the ambient gas, we measured the pressure jump at the discontinuity. Using the radial surface brightness distribution in Figure 8 and the temperature distribution in Figure 9, we determine a 3-D density profile, applying the deprojection technique given by Kriss et al. (1983), but taking into account the temperature gradient. We find that the density increases by a factor of $4.1 \pm 0.6$ across the discontinuity, from $3.2 \times 10^{-4}$ cm$^{-3}$ to $1.3 \times 10^{-3}$ cm$^{-3}$. Multiplying by the temperature drop from 1.1 to 0.7 keV (Figure 9), we measure a factor of 2.6 pressure jump. Following the formula in Markevitch & Vikhlinin (2007), the pressure jump corresponds to a Mach number, $M$ =1.2. If the metal abundance also changes across the discontinuity, as seen in the section 4 from sub-solar to super-solar, the density jump will be considerably reduced because the X-ray emissivity depends on the metal abundance. Taking the abundance change from 0.5 to 1.5 solar into account, we re-measure the pressure jump to be 1.8, which in turn corresponds to the Mach number of 0.9. With the sound speed $C_S$ = 540 km s$^{-1}$ in the ambient gas (at 1.1 keV), the galaxy velocity relative to the ICM is 480 - 650 km s$^{-1}$, depending on the metal abundance change across the discontinuity (the lower velocity is more likely, given the abundance difference between the ISM and ICM). The radial velocity (3820 km s$^{-1}$) of NGC 7619 is ~300 km s$^{-1}$ higher than the mean radial velocity (3525 km s$^{-1}$) of the group galaxies (Ramella et al. 2002). If the excess radial velocity of NGC 7619 represents the radial motion relative to the ambient gas, the velocity vector of NGC 7619 is at 25-32° from the sky plane.

Similar motions were first reported in clusters of galaxies (e.g., A3667, Vikhlinin et al. 2001), where the pressure jump at the cold front corresponds to the Mach number close to 1, or the velocity ~ 1200 - 1600 km s$^{-1}$ and in a few elliptical galaxies by analyzing the cold fronts identified with recent *Chandra* observations. Machacek et al. (2005) derived $M$ = 0.8-1.0 (or $V$ = 530-660 km s$^{-1}$) in NGC 1404, an elliptical galaxy close to NGC 1399 in the Fornax cluster, but the tail in NGC 1404 is not as extended as in NGC 7619. In NGC 4552, an elliptical galaxy in the Virgo cluster, Machacek et al. (2006) derived a supersonic motion with $M$ = 1.9-2.7 (or, $V$ = 1460 - 2070 km s$^{-1}$). In NGC 4552, the tail is extended to ~10 kpc to the south while the cold front is seen at ~3 kpc north of the galaxy.

The possible causes of the head-tail structure and the cold front are ram-pressure stripping (as suggested in NGC 1404 and NGC 4552) and sloshing (as suggested in the center of relaxed clusters; see Markevitch & Vikhlinin 2007 for more examples and a general review on these subjects). We consider both mechanisms and discuss their applicability to the observed head-tail structure in NGC 7619. While the galaxy orbits around inside the hotter ambient medium, the galaxy ISM will experience the ram-pressure and will form a head-tail structure during the stripping process. This has also been identified as one of major mechanisms of the ICM metal enrichment (Gunn & Gott 1972). While the head-tail structure in NGC 7619 looks similar to those in other galaxies experiencing ram-pressure stripping, it is not easy to understand how this galaxy moves around relative to the ambient hotter medium. This is because NGC 7619 appears to sit at the center of the Pegasus I group/cluster, or at one of the two local potential minima, unlike NGC 1404 (10′ away from the dominant galaxy in Fornax cluster, NGC 1399) and 4552 (1° away from the dominant galaxy in Virgo cluster, M87). NGC 7619 and NGC 7626 (7′ apart) are the two biggest galaxies in this group. Both of them are ellipticals and are almost identical in their optical size and luminosity. However, NGC 7619 may move through or oscillate near the bottom of the potential well, as seen in the A1795 cD galaxy (Fabian et al. 2001). Or, NGC 7619 may be bound in a binary system with NGC 7626 and is possibly moving on the binary orbit. The projected distance between NGC 7619 and NGC 7626 is only 106 kpc. The motion of NGC 7619 relative to the ambient gas is also consistent with the radial velocity of NGC 7619 being one of the highest among the group galaxies, ~300 km s$^{-1}$ higher than the mean radial velocity (Ramella et al. 2002).

On the other hand, the sloshing mechanism may be a better explanation for the head-tail structure in NGC 7619 than ram-pressure stripping, if the galaxy is not moving at the bottom of the potential well. Sloshing was intended to explain the cold front often seen in the cores of relaxed, cooling flow clusters (e.g., Ascasibar & Markevitch 2006). Sloshing of cold gas in the central gravitational potential which might be initiated by the perturbation due to minor mergers (by a gasless sub-cluster) may reproduce observational features of smooth discontinuity often seen in the cold fronts without any other obvious disturbance which might be caused by major mergers (see Markevitch & Vikhlinin 2007 for more details). However, sloshing requires a steep entropy gradient (as in cooling flow clusters, Ascasibar & Markevitch 2006), while the entropy gradient is relatively small in NGC 7619 due to a smaller temperature change than seen in cooling flow clusters. The sloshing of the cold gas often produces multiple cold fronts (or a spiral structure) due to the repeated Rayleigh-Taylor like instability against ram-pressure from the surrounding hot gas. However, we do not see multiple discontinuities in NGC 7619, although it is still possible that other cold fronts near the center may be hidden by the inclination effect (~30°). While NGC 7626 may be an excellent candidate for the initial perturbation which is necessary for the sloshing, NGC 7626 also contains a significant amount of the hot ISM (section 4.5). The passage of NGC 7626 would have made the X-ray surface brightness of NGC 7619 much more complex as seen in the merger simulation with a gas-rich sub-cluster (Ascasibar & Markevitch 2006).

Out of the two possible causes of the head-tail structure in NGC 7619, the morphological structure seems to support ram-pressure stripping, while the fact that NGC 7619 is sitting at or near the center of the Pegasus I group/cluster prefers sloshing. We consider that our data do not allow us to exclusively select one mechanism over another. However, we note that regardless of the origin of the head-tail structure (either ram-pressure stripping or sloshing), the physical status of the hot ISM remains the same, as the hot ISM experiences the pressure from the NW direction.

Another possibility for the extended tail structure may be tidal interaction with the nearby galaxy (e.g., NGC 7626). Observational evidences for tidal interaction between NGC 7619 and 7626 have been reported (though mostly in NGC 7626), including a kinematically peculiar core in both NGC 7619 and NGC 7626 (Bender 1990; Balcells and Carter 1993) and an optical excess feature (after subtracting a smooth galaxy model) of NGC 7626 toward NGC 7619 (Forbes and Thomson 1992). However, it is hard to explain the sharp discontinuity at the leading edge by tidal interaction only.

5.2 Metal abundances in the ISM and the surrounding gas

Heavy elements in the hot ISM of elliptical galaxies are the relics of stellar evolution. The stellar evolution models of elliptical galaxies predict that the metallicity in the hot ISM is higher than (or at least as high as) that observed in the stellar system, i.e., super-solar metal abundance ($Z_{Fe}$ = 2-5 times solar, Arimoto et al. 1997; ~10 times solar, Pipino et al. 2005). Iron in the hot ISM, which exhibits the strongest X-ray emission features, is expected to be at least similar to (or higher than) that of the stellar population in elliptical galaxies, where Iron was initially synthesized by the bulk of Type II supernova (*SN*) explosions and then enriched during the lifetime of the galaxy by Type Ia *SN*. We have measured the *Fe* abundance to be super-solar (1-2 times solar) within the $D_{25}$ ellipse of NGC 7619. This is consistent with the theoretical expectation from the accumulation of *SN* synthesized metals in this galaxy and those of other bright elliptical galaxies (e.g., NGC 507 Kim and Fabbiano 2004; NGC 1399 Buote 2002; NGC 5044 Buote et al. 2003).

At the outskirts (i.e., outside the head-tail structure), the metal abundance is ~1/2 $Z_\odot$, close to that in the typical ICM (e.g., Mushotzky et al. 1996; Fukazawa et al. 1998) and the temperature (1.1 keV) is hotter than that (0.7 keV) of the ISM. On the other hand, the gas in the X-ray tail is similar to the hot ISM within the $D_{25}$ ellipse both in the metal abundance (super-solar) and temperature (0.8 keV), but quite different from the ambient gas, indicating that the X-ray tail is indeed originated from the galaxy. Taking the mass loss rate from Faber & Gallagher (1976), we estimate the accumulated mass from stellar mass loss over the Hubble time to be $1.5 \times 10^{10}$ $M_\odot$, which is comparable to the total mass of the hot gas in the core-tail structure ($2 \times 10^{10}$ $M_\odot$). Given that the mass loss rate would have been higher in the past when the star formation rate was higher, the stellar mass loss would be enough to explain the total hot gas in the core-tail structure.

Determining the relative abundance of *Fe* and α-elements is critical for discriminating between the relative importance of *SN* type II and type Ia in the parent galaxy (e.g., Renzini et al. 1993; Loewenstein et al. 1994). Therefore, these measurements provide important clues for our understanding of the evolution of both stellar component and hot ISM. If heavy elements are mainly synthesized in Type II *SNe*, the abundance ratio of α-elements to *Fe* is expected to be higher than the solar ratio (e.g., Woosley et al. 1995), while the ratio decreases with increasing contribution from Type Ia *SNe* (e.g., Iwamoto et al. 1999). In section 4, we show that the abundance ratio of α-elements to *Fe* is close to (or slightly lower than) the solar ratio. With SN yields taken from Gibson et al. (1997) and converted to the revised solar values given by Grevesse and Sauval (1998), the measured abundance ratio of *Si/Fe* (near solar) indicates that 60-80% of the detected Iron mass is produced in SN Type Ia.

The ram-pressure stripping in the dense environment, as one of the important ICM metal enrichment mechanisms, will remove the ISM with heavy elements synthesized via stellar evolution and spread out into the surrounding ICM. If the head-tail structure is formed by ram-pressure stripping (see section 5.1), the extended material coming from the galaxy can provide the direct evidence of the ICM metal enrichment by on-going ram pressure stripping process. On the other hand, sloshing would be less efficient, because the bulk of ISM is still bound to the galaxy. Other possible mechanisms for the ICM metal enrichment are a galactic wind (e.g., de Young 1978), galaxy-galaxy interaction (e.g., Kapferer et al. 2005) and inter-cluster supernovae (Domainko et al. 2004). One of the key measurements to distinguish various mechanisms is to determine the α-element to *Fe* abundance ratio, because different types of *SNe* preferentially produce different elements. For example, the galactic wind driven by *SNe* type II during the early star formation period could remove the α-element enhanced ISM from the galaxy and inject them to the ICM. Unfortunately, we can only loosely constrain the abundance ratio (being close to solar) inside the $D_{25}$ ellipse, but can not tightly constrain it in the X-ray tail or in the ambient gas.

The best-fit hydrogen column ($N_H$ = 1.2 x $10^{21}$ cm$^{-2}$ for MOS and 9 x $10^{20}$ cm$^{-2}$ for PN; see Table 4) in the center of NGC 7619 is slightly higher than the Galactic line of sight value (5.0 x $10^{20}$ cm$^{-2}$), while it becomes lower (2 x$10^{20}$ cm$^{-2}$) than the Galactic value in the outer region (r=3′-5′). Neither IRAS FIR (Knapp et al. 1989) nor HI observations (Knapp et al. 1985) of NGC 7619 indicate significant internal absorption, yielding an upper limit of $M_{HI}$ of ~8 x $10^8$ M$_\odot$ (after correcting for the different distance). Since an intrinsic hydrogen column of a few x $10^{20}$ cm$^{-2}$ within the $D_{25}$ ellipse of NGC 7619 would correspond to $M_{HI}$ = a few x $10^9$ M$_\odot$, we can rule out the presence of internal absorption in NGC 7619, unless there is a significant amount of molecular gas (see Arabadjis and Bregman 1999). We note that $N_H$ and the amount of hard component returned by the spectral fits are partially tied, in the sense that a larger $N_H$ tends to go with a smaller hard component. This in turns would affect the model predictions for the thermal continuum at low (E < 0.7 keV) and high energies (E > 2 keV), slightly reducing the required strength of the *Fe* peak at ~1 keV and hence the *Fe* abundance (see also Kim & Fabbiano 2004).

6. CONCLUSIONS

Analyzing images and spectra of the head-tail structure in the hot ISM of NGC 7619 obtained with *Chandra* ACIS and *XMM-Newton* observations, we conclude the following:

(1) The hot ISM (0.7 keV) of NGC 7619 consists of a long extended tail to the SE direction and a discontinuity in the leading edge to the opposite direction. The jump condition at the discontinuity suggests that NGC 7619 is moving to the NE direction at a velocity of ~500 km s$^{-1}$, against the ram-pressure imposed by the hotter (1.1 keV) ambient gas.

(2) The *Fe* abundance within the $D_{25}$ ellipse of NGC 7619 is super solar (1-2 times solar). This is consistent with the theoretical expectation from the stellar evolution models and those recently measured in other X-ray bright elliptical galaxies. Instead, the *Fe* abundance in the outskirts is sub solar (≤ 0.5 solar), consistent with the typical ICM abundance.

(3) The *Fe* abundance at the extended tail is enriched (super solar) and higher than that in the surrounding region. The temperature of the tail is also closer to the cooler ISM than to the hotter ICM, indicating that the gas in the X-ray tail is originated from NGC 7619. The possible cause of the head-tail structure in NGC 7619 is either on-going ram pressure stripping or sloshing.

(4) The X-ray spectra of NGC 7626 show that the hot ISM (0.7 keV with solar metallicity) and LMXBs (or possibly AGN) contribute 60 and 40% of the total X-ray emission. While NGC 7617 and NGC 7623 are similar to NGC 7626 in having both soft gas and hard LMXB components, NGC 7611 seems to contain no or little gas.

(5) One off-nuclear point source is detected within the $D_{25}$ ellipse of NGC 7619. If it is in the galaxy, it is a typical ULX with $L_X = 5 \times 10^{39}$ erg s$^{-1}$. The X-ray emission of the ULX candidate is soft, possibly suggesting that it is not a background AGN.

This work was supported by NASA grant G03-4109X and NNGO4GC63G. We thank the anonymous referee and M. Markevitch for helping us to improve the discussion on the sloshing mechanism.

**Figure Captions**

Figure 1. The field of view of X-ray observations overlaid on the DSS optical image. The XMM MOS observation is indicated by a blue circle, while the ACIS-S observation by green squares and the ACIS-I observation by red squares. Detected X-ray sources are marked by circles (PSF sizes) with the same color as used for each fov. Also labeled are four galaxies (including our target, NGC 7619) detected in these X-ray observations.

Figure 2. Gaussian-smoothed (σ=5″) soft-band (0.3-2.5 keV) ACIS S3 image with the $D_{25}$ ellipse (large blue ellipse) and detected sources (small red circle) marked.

Figure 3. Gaussian-smoothed (σ=7.5″) soft-band (0.3-2.5 keV) *XMM-Newton* image (MOS 1+2 combined) with the $D_{25}$ ellipse (large blue ellipse) and detected sources (small red circle) marked.

Figure 4. The number of point sources in the X-ray tail is compared with the cosmic background log(N)-log(S) taken from the ChaMP data (Kim, M. et al. 2007). The sources in the hard band (triangle) should be compared with the hard band ChaMP prediction (dashed line); sources in the soft band (circles) with the soft band prediction (solid line).

Figure 5. The exposure-corrected, point-source-excluded, Gaussian-smoothed (σ=12″), narrow-band (0.7-1.2 keV) image.

Figure 6. Same as Figure 5, but head and tail regions marked. The green and magenta pies indicate the head and tail directions, respectively. The blue ellipse roughly indicates the X-ray surface brightness distribution of the hot ISM. The red ellipses indicate the tail regions where the spectra were extracted.

Figure 7. Radial profiles of X-ray surface brightness in 0.7-1.2 keV toward the NE (head in red) and SW (tail in black) directions, determined with the ACIS S3 (filled) and the MOS 1+2 data (open).

Figure 8. Same as Figure 7, but zoom-in view near the discontinuity in a linear distance scale.

Figure 9. Emission temperature against radius toward the NE direction (the head side in PA= -20° — 100°).

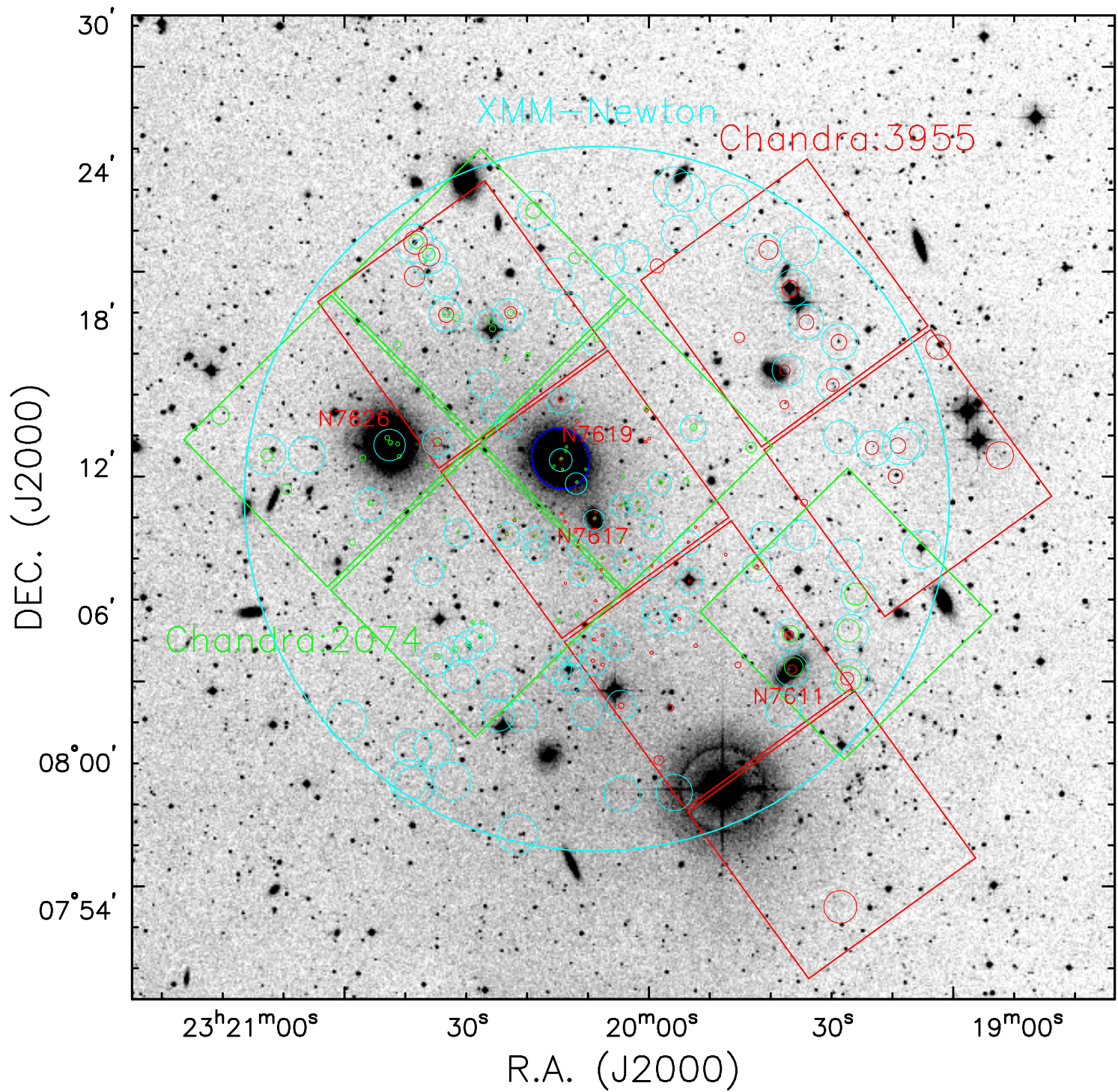

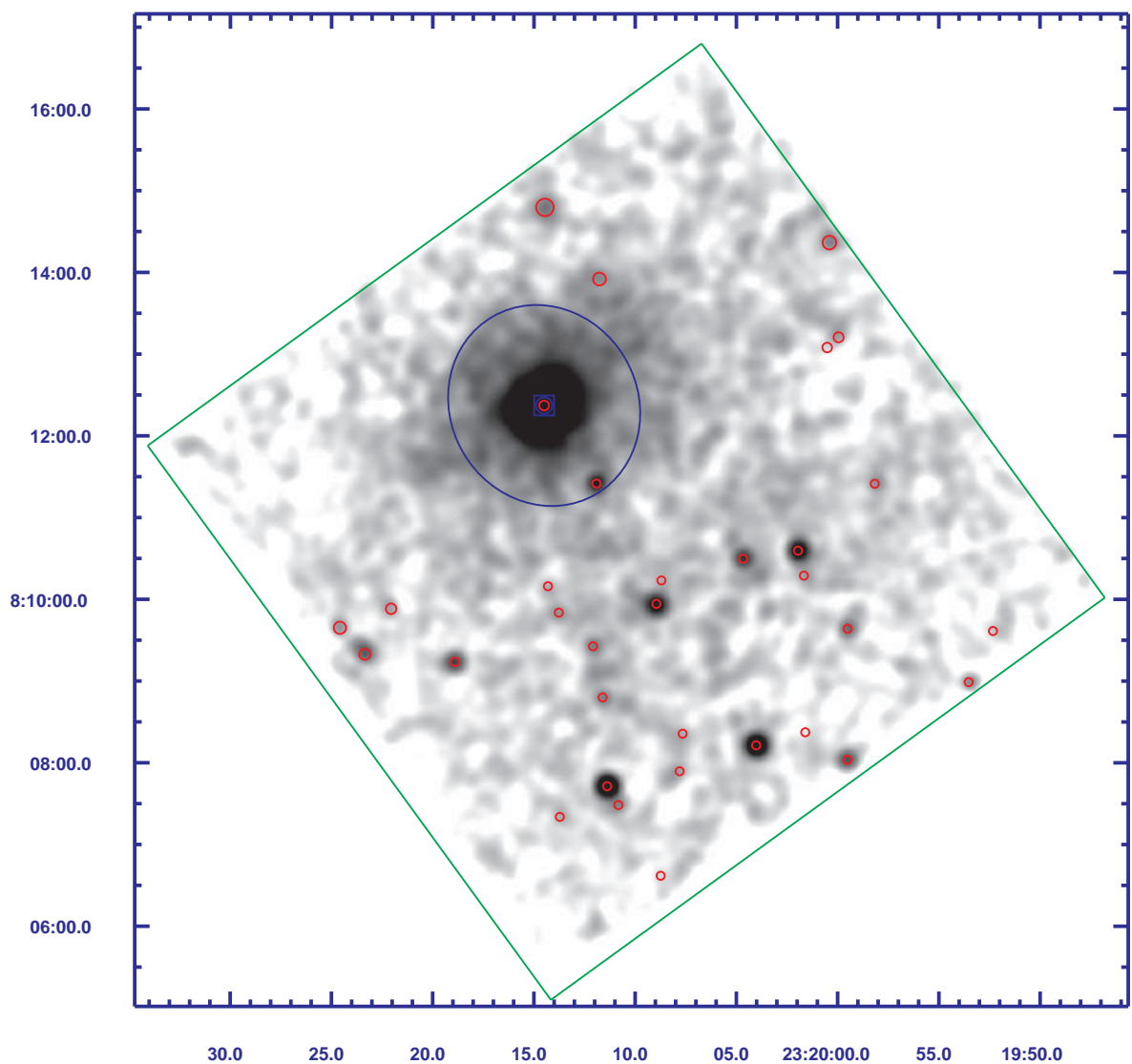

NGC 7619

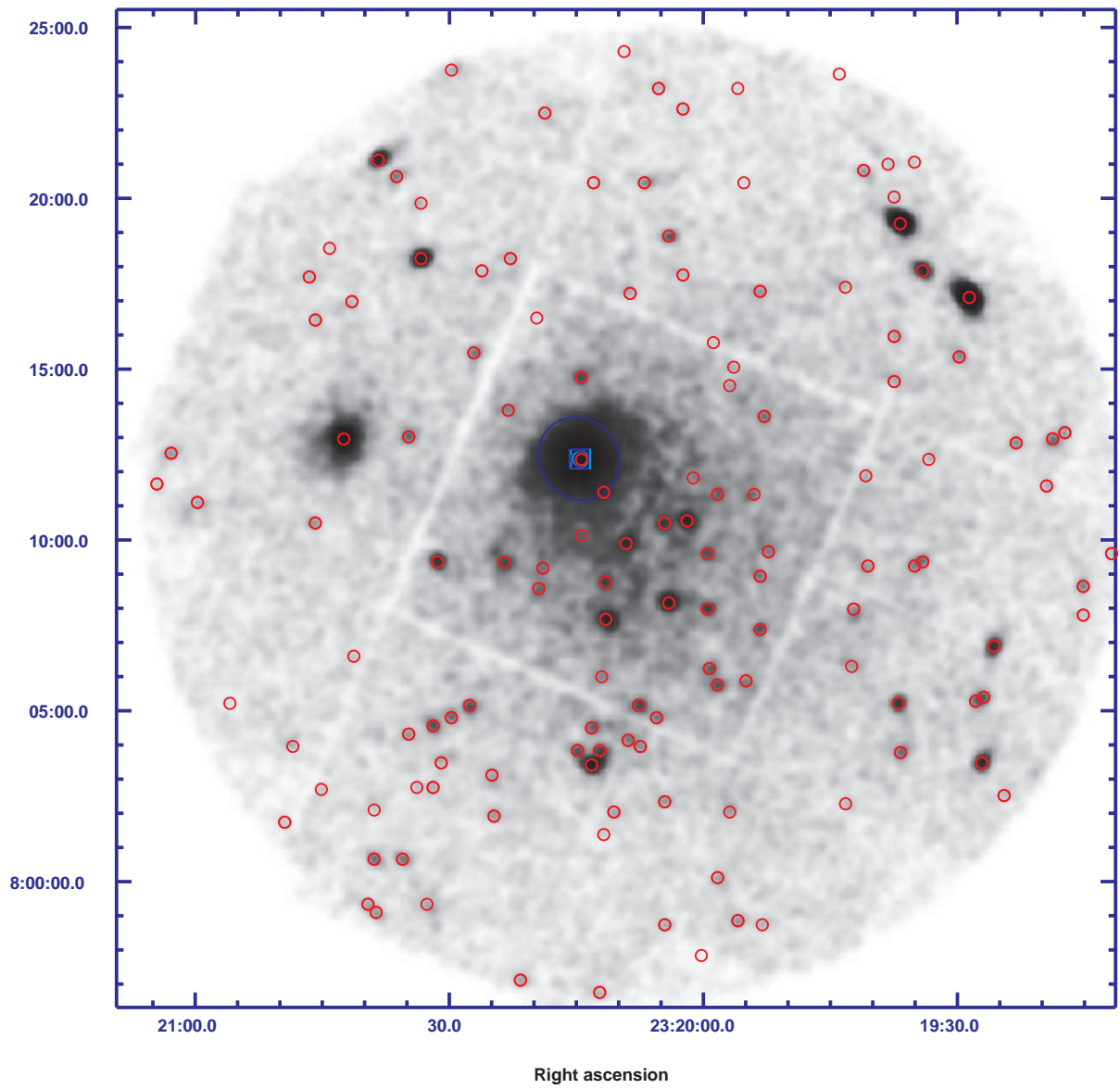

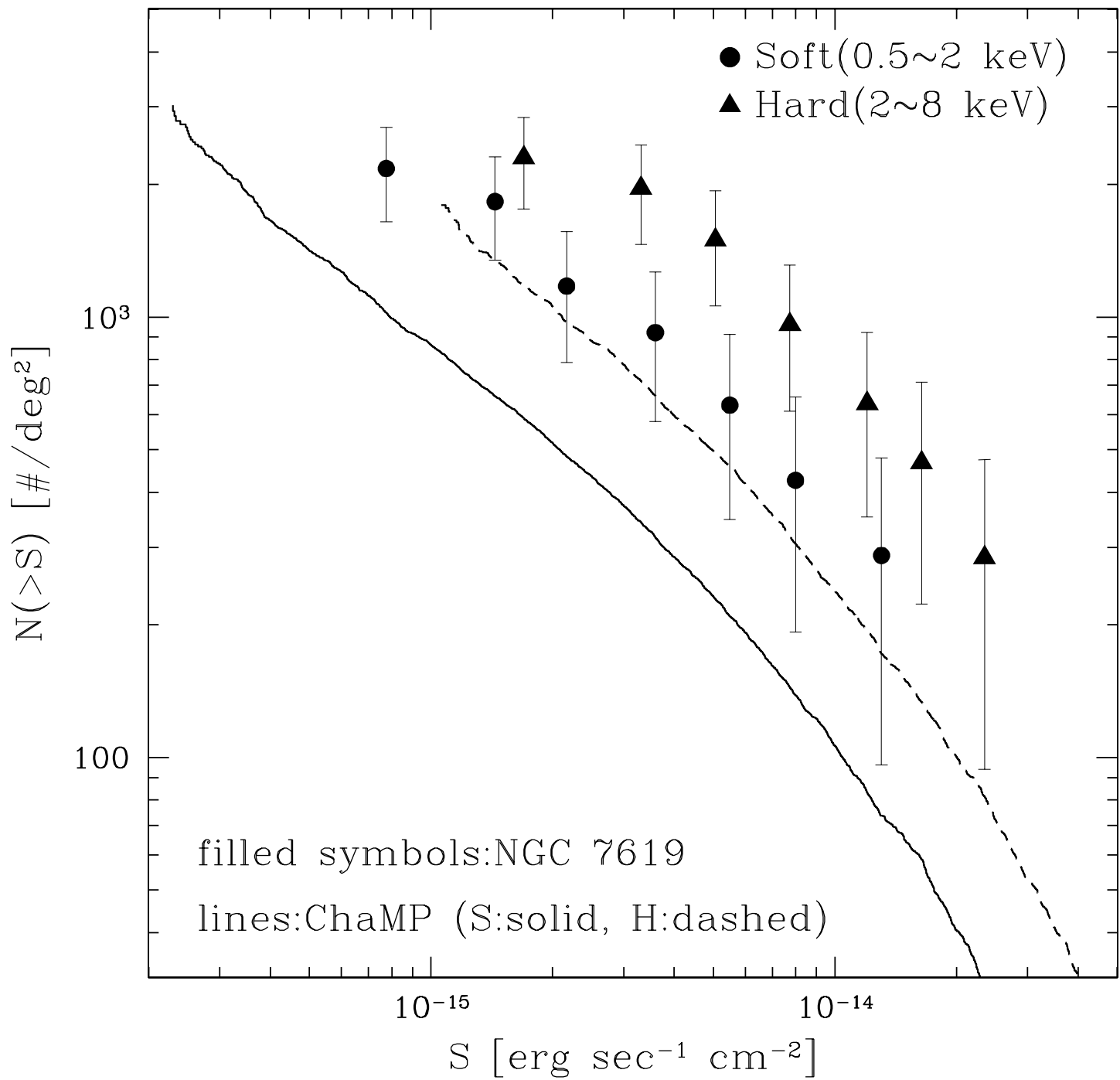

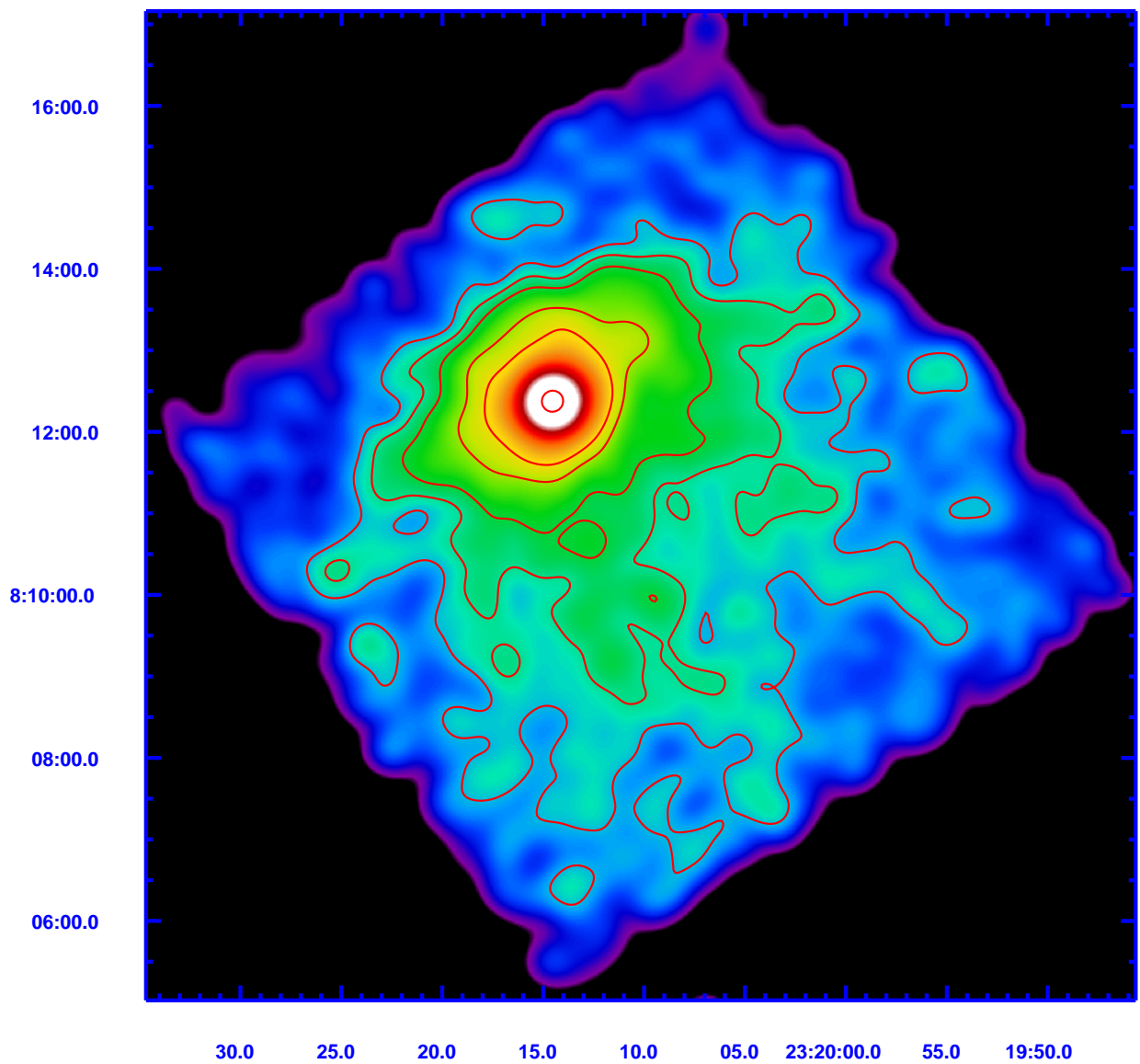

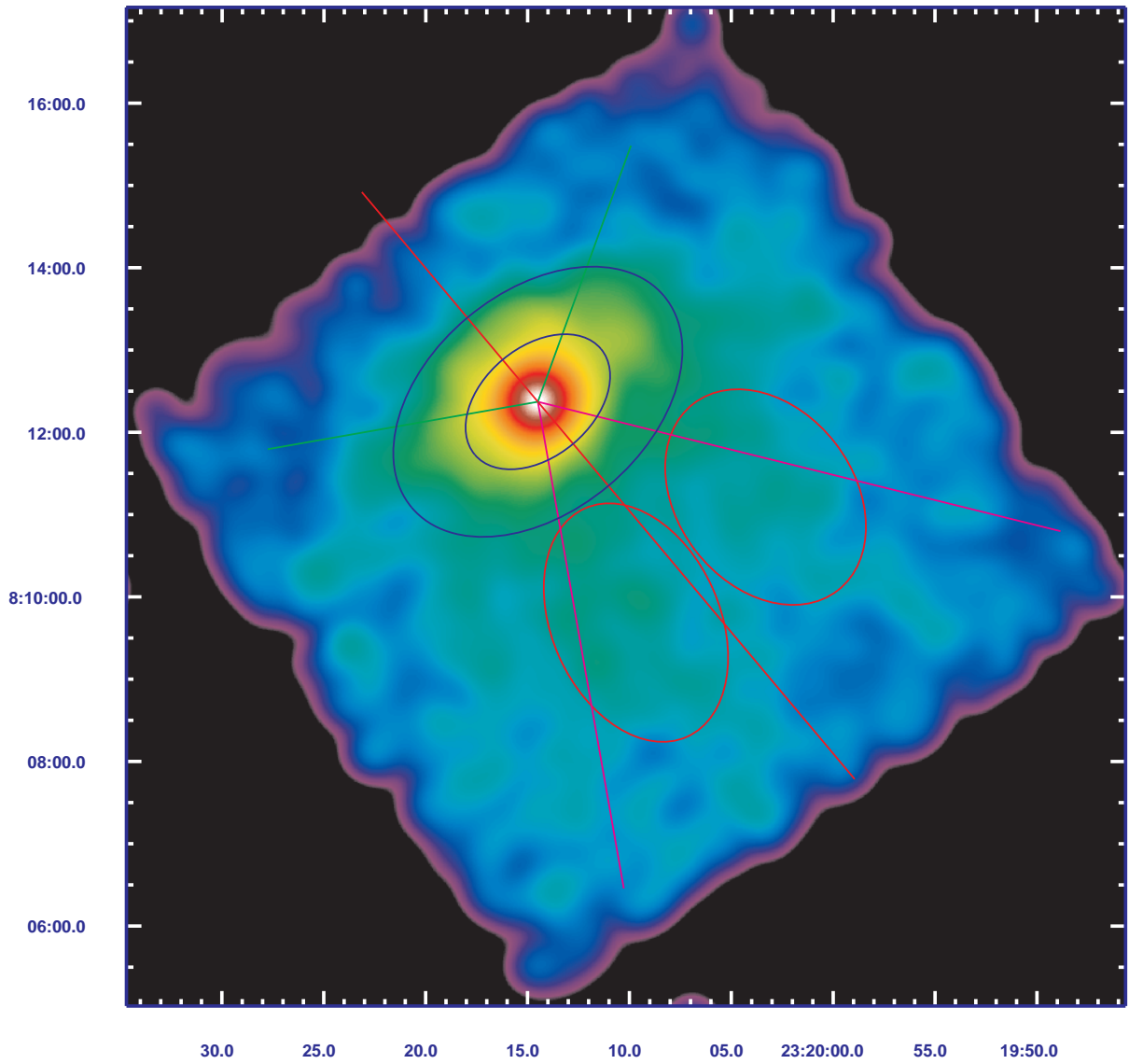

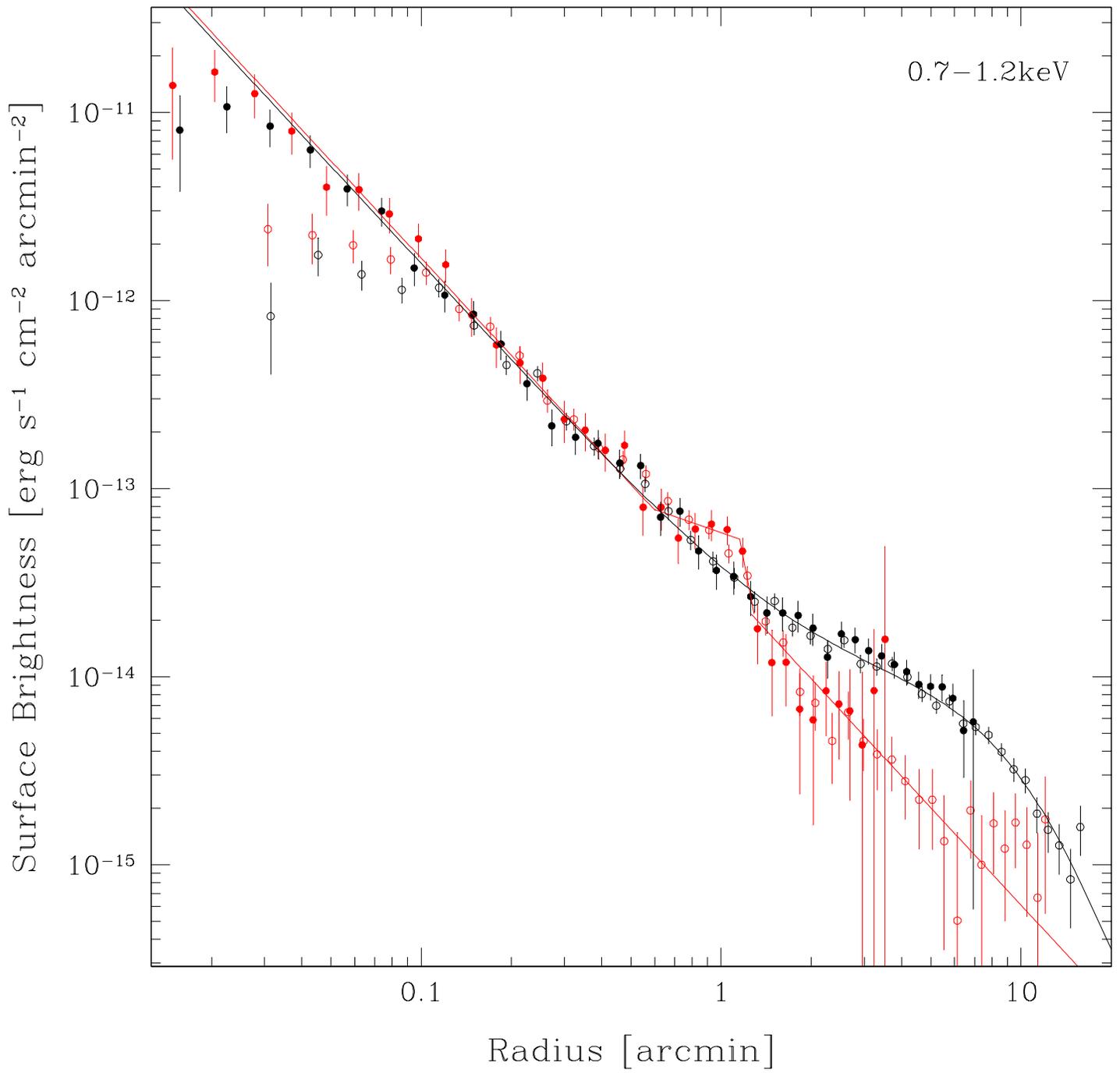

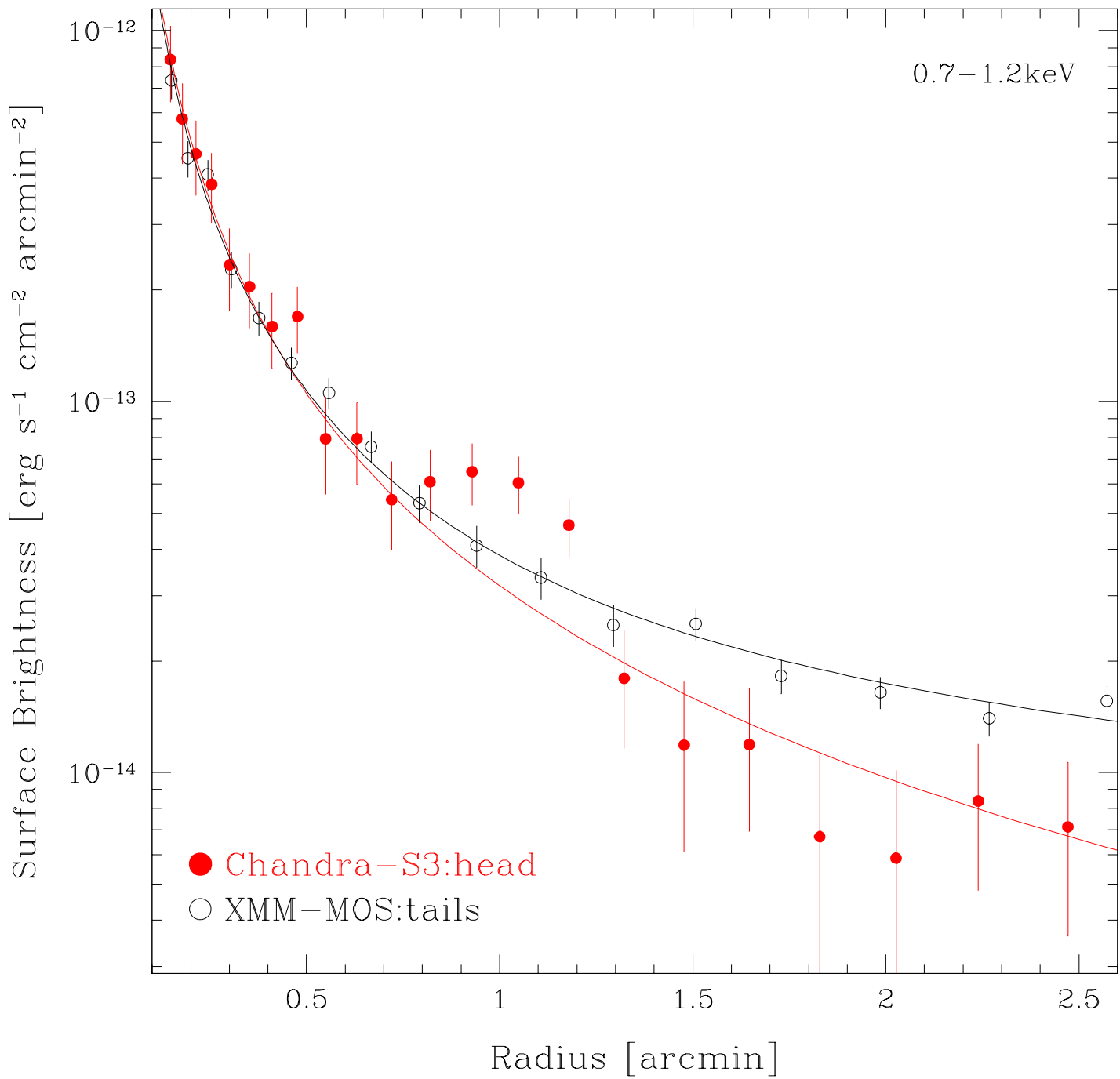

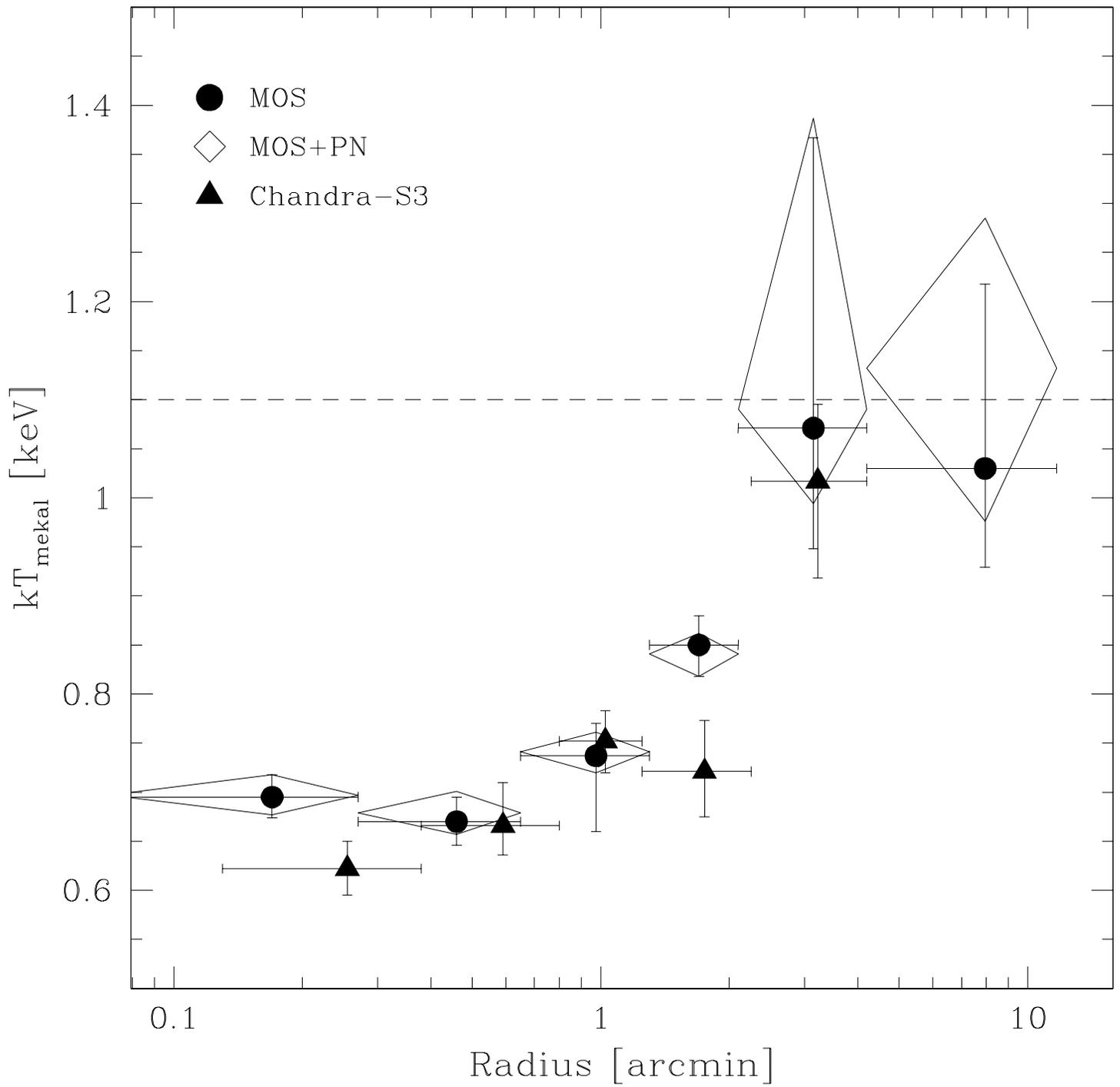